\documentstyle{article}

\makeatletter
\newcommand{\noepsf}{\protect\newcommand{\insertfigs}[2]{}%
  \newdimen\epsfxsize%
  \protect\newcommand{\epsffile}[1]{\begin{center}{\Large\em Figure goes
here.}\end{center}}}
\def\answy{y }
\def\answl{l }
\def\interactive{\def\intdone{}\typeout{}%
\typeout{This is interactive mode. If it annoys you, comment out the}
\typeout{"interactive" command near the beginning of the document.} \typeout{}
\ifx\nofigures\undefined\typeout{This paper should have come with seperate
postscript figure(s).} \typeout{Including them in the text requires the file
epsf.tex in addition} \typeout{to the figures themselves.}%
\message{Include figures? (y/n) }\read-1 to\answ%
\ifx\answ\answy
  \input{epsf.tex}
  \ifx\epsffile\undefined\typeout{Unable to load epsf.tex. Figures will *not*
be included.}\typeout{}
  \noepsf
  \else\typeout{epsf.tex loaded. Figures will be included.}\typeout{}
  \fi
\else
  \typeout{Figures will not be included.}\typeout{}
  \noepsf
\fi
\fi
\typeout{Portrait (large, one-up) or landscape (small, two-up) mode?}
\message{(If confused, enter p.) (p/l) }%
\read-1 to\answ%
\ifx\answ\answl
  \landscapemode
  \typeout{Paper will be produced in landscape mode.}\typeout{}
\else
  \makeatletter\input{art12.sty}
	\renewcommand{\baselinestretch}{1.1}
	\textwidth=6.5in
	\textheight=9.0in
	\hoffset=-0.50in
	\voffset=-1.0in
	\def\draftstamp{\draftstampsubmit}
	\def\TitlePageNoNumber{\thispagestyle{empty}}
  \typeout{Paper will be produced in portrait mode.}\typeout{}
\fi
}
\def\printpagenum{\ifnum\thepage=0 \else \thepage \fi}
\def\landscapemode{%
	\renewcommand{\baselinestretch}{1.1}
	\twocolumn\sloppy\flushbottom\parindent 2em
	\leftmargini 2em\leftmarginv .5em\leftmarginvi .5em
	\oddsidemargin 0in	\evensidemargin 0in
	\columnsep .4in	\footheight 0pt
	\textwidth 10in	\topmargin  -.4in
	\headheight 0pt \topskip 0in
	\textheight 6.9in \footskip 30pt
	\hoffset -.5in \voffset -.25in
	\def\@oddfoot{\hfil\printpagenum\hfil\addtocounter{page}{1}
		\hspace{\columnsep}\hfil\printpagenum\hfil}
	\let\@evenfoot\@oddfoot	\def\@oddhead{}	\def\@evenhead{}
	\def\draftstamp{\draftstamptwoup}
	\def\TitlePageNoNumber{}}

\def\draftstampsubmit{
}

\def\draftstamptwoup{
}

    \def\eqnarray{\stepcounter{equation}\let\@currentlabel=\theequation
    \global\@eqnswtrue
    \global\@eqcnt\z@\tabskip\@centering\let\\=\@eqncr
    $$\halign to \displaywidth\bgroup\@eqnsel\hskip\@centering
      $\displaystyle\tabskip\z@{##}$&\global\@eqcnt\@ne
       \hfil${{}##{}}$\hfil
      &\global\@eqcnt\tw@ $\displaystyle\tabskip\z@{##}$\hfil
       \tabskip\@centering&\llap{##}\tabskip\z@\cr}

\def\section{\@startsection{section}{1}{\z@}{-1.5ex}{1.5ex}%
{\large\bf}}
\def\subsection{\@startsection{subsection}{2}{\z@}{-1.2ex}{1.2ex}%
{\normalsize\bf}}
\def\subsubsection{\@startsection{subsubsection}{3}{\z@}{-1ex}{1ex}%
{\small\bf}}

\def\preprintnum#1{\hspace*{\fill #1}\par}
\def\preprintdate#1{\hspace*{\fill #1}\par}
\def\title#1{\TitlePageNoNumber\vskip2.5pc
    \begin{center}
    {\LARGE #1 \par}%
    \end{center}
    \vskip 1.5em}
\def\author#1#2{\renewcommand{\thefootnote}{\fnsymbol{footnote}}
	\begin{center}
    {\large \lineskip .5em
    #1\footnote{e-mail: {\tt #2}} \par}
    \end{center}
    }
\def\institution#1{\begin{center}
    {\large \lineskip .5em \em #1 \par}
    \end{center}
    \vskip 1.5em}
\def\abstract#1{{\small
    \begin{center}%
    {\bf Abstract\vspace{-.5em}\vspace{\z@}}%
    \end{center}%
    \quotation #1 \endquotation\newpage}}

\def\be#1{\begin{equation} \label{#1}}
\def\ee{\end{equation}}
\def\bes#1{\begin{eqnarray} \label{#1}}
\def\ees{\end{eqnarray}}
\def\nn{\nonumber}

\newlength{\minuswidth}
\newlength{\digitwidth}
\newlength{\pointwidth}
\newcommand{\am}{\hspace*{\minuswidth}} 
\newcommand{\ad}{\hspace*{\digitwidth}} 
\newcommand{\ap}{\hspace*{\pointwidth}} 

\def\citen#1{%
\edef\@tempa{\@ignspaftercomma,#1, \@end, }
\edef\@tempa{\expandafter\@ignendcommas\@tempa\@end}%
\if@filesw \immediate \write \@auxout {\string \citation {\@tempa}}\fi
\@tempcntb\m@ne \let\@h@ld\relax \let\@citea\@empty
\@for \@citeb:=\@tempa\do {\@cmpresscites}%
\@h@ld}
\def\@ignspaftercomma#1, {\ifx\@end#1\@empty\else
   #1,\expandafter\@ignspaftercomma\fi}
\def\@ignendcommas,#1,\@end{#1}
\def\@cmpresscites{%
 \expandafter\let \expandafter\@B@citeB \csname b@\@citeb \endcsname
 \ifx\@B@citeB\relax 
    \@h@ld\@citea\@tempcntb\m@ne{\bf ?}%
    \@warning {Citation `\@citeb ' on page \thepage \space undefined}%
 \else
    \@tempcnta\@tempcntb \advance\@tempcnta\@ne
    \setbox\z@\hbox\bgroup 
    \ifnum\z@<0\@B@citeB \relax
       \egroup \@tempcntb\@B@citeB \relax
       \else \egroup \@tempcntb\m@ne \fi
    \ifnum\@tempcnta=\@tempcntb 
       \ifx\@h@ld\relax 
          \edef \@h@ld{\@citea\@B@citeB}%
       \else 
          \edef\@h@ld{\hbox{--}\penalty\@highpenalty \@B@citeB}%
       \fi
    \else   
       \@h@ld \@citea \@B@citeB \let\@h@ld\relax
 \fi\fi%
 \let\@citea\@citepunct
}
\def\@citepunct{,\penalty\@highpenalty\hskip.13em plus.1em minus.1em}%
\def\@citex[#1]#2{\@cite{\citen{#2}}{#1}}%
\def\@cite#1#2{\leavevmode\unskip
  \ifnum\lastpenalty=\z@ \penalty\@highpenalty \fi 
  \ [{\multiply\@highpenalty 3 #1
      \if@tempswa,\penalty\@highpenalty\ #2\fi 
    }]\spacefactor\@m}
\renewcommand{\cite}[1]{[\citen{#1}]}

\def\slashchar#1{\setbox0=\hbox{$#1$}           
   \dimen0=\wd0                                 
   \setbox1=\hbox{/} \dimen1=\wd1               
   \ifdim\dimen0>\dimen1                        
      \rlap{\hbox to \dimen0{\hfil/\hfil}}      
      #1                                        
   \else                                        
      \rlap{\hbox to \dimen1{\hfil$#1$\hfil}}   
      /                                         
   \fi}                                         %
\newcommand{\bra}[1]{\mbox{$\left\langle{#1}\right|$}}
\newcommand{\ket}[1]{\mbox{$\left|{#1}\right\rangle$}}

\newcommand{\tr}{\mbox{Tr}}
\newcommand{\bold}[1]{\mbox{\boldmath$#1$}}

\newcommand{\eqn}[1]{Eq.~(\protect\ref{#1})}

\newcommand{\fig}[1]{Fig.~\protect\ref{#1}}
\newcommand{\tabref}[1]{Table~\protect\ref{#1}}

\newcommand{\secref}[1]{Section~\protect\ref{#1}}

\newcommand{\sw}[1]{{\mbox{\scriptsize #1}}}

\interactive

\newcommand{\vecp}{\bold{p}}
\newcommand{\vecr}{\bold{r}}
\newcommand{\vecL}{\bold{L}}
\newcommand{\lambar}{\overline{\Lambda}}
\newcommand{\gtlt}{\begin{array}{c}>\\[-10pt]<\end{array}}

\begin{document}

\preprintnum{EFI-93-39}
\preprintnum{hep-th/9309270}
\preprintdate{August, 1993}

\title{Subleading Heavy Quark Effects in a Nonrelativistic Quark Model}

\author{James F. Amundson\footnote{Address after Sept.\ 1, 1993: {\em
Department of Physics, University of Wisconsin, Madison, WI 53706}}}%
{amundson@yukawa.uchicago.edu}

\institution{Enrico Fermi Institute and Department of Physics,\\
University of Chicago,\\
5640 S. Ellis Ave., Chicago, IL 60637}

\abstract{I present a very simple model, based on the model of Isgur,
Scora, Grinstein and Wise \cite{ISGW}, which can be used to calculate the
effects which appear at subleading order in heavy quark effective theory. I
include both general formalism and specific results. The formalism
transparently reproduces the results of heavy quark effective theory, while
giving insight into such things as the vanishing of certain form factors at
zero recoil. I discuss the implications of these results for both heavy
quark effective theory and the Isgur-Scora-Grinstein-Wise model.}

\section{Introduction}

The discovery of heavy quark symmetry and the subsequent formulation of
heavy quark effective theory (HQET)
\cite{VolShif,NussWetz,PolWise,IsWise,EichHill,Georgi,GeorgiReview,%
GrinsteinReview,IsWiseReview,MannelReview,NeubertReview} have substantially
improved the theoretical understanding of hadronic systems involving heavy
quarks, i.e., quarks with $m_Q\gg\Lambda_\sw{QCD}$. Before HQET,
calculations involving heavy hadrons involved models such as QCD sum rules,
constituent quark models, etc. Models such as these inevitably rely on
various assumptions, and the errors due to these assumptions are difficult
to quantify.

HQET, however, provides a systematic expansion of the QCD Lagrangian in
powers of $\lambar/m_Q$, where $\lambar$ is a constant of order
$\Lambda_\sw{QCD}$. It also provides an unambiguous procedure for
incorporating radiative corrections. We can therefore argue that a
calculation in the effective theory to a given order in $\lambar/m_Q$ and
$\alpha_s$ is complete up to corrections that are higher order in these
small parameters.

The price we pay for this confidence in calculations is predictive power.
As an example, at leading order in $\lambar/m_Q$, the HQET predictions for
form factors in $B\to D^{(*)}l\nu$ decay involve one unknown function,
$\xi(w)$, known as the Isgur-Wise function. The Isgur-Wise function is not
calculable in perturbative QCD. (It is, however, calculable in principle
using lattice gauge theory.) Heavy quark symmetry tells us the
normalization of the Isgur-Wise function at the zero-recoil point, i.e.,
the point where the initial and final heavy quarks have the same
four-velocity. Although the radiative corrections can be simply calculated
using perturbative QCD, the $1/m_c$ corrections induce new uncalculable
functions. At subleading order in $\lambar/m_c$ for $B\to D^{(*)}l\nu$
decays, there are four new uncalculable functions \cite{LukeThm}. This time
the normalizations are not determined, but two of the functions have been
shown to vanish at the zero-recoil point \cite{LukeThm}. Since there are
only four measurable form factors in $B\to D^{(*)}l\nu$ decay (there are
six if the mass of the lepton $l$ is not neglected), the predictive power
of the theory is dramatically reduced. Fortunately, the vanishing of the
subleading functions is such that the prediction for the differential rate
at zero recoil in $B\to D^{*}l\nu$ decay is determined to ${\cal
O}(\lambar^2/m_c^2)$. This result has been used to obtain a
model-independent extraction of $V_{cb}$ \cite{NeubertReview,NeubertVcb}
with an error that is currently limited by experiment, not theory. Aside
from a few specific cases such as this, however, there is very little
predictive power in the theory at subleading order. At higher orders in
$\lambar/m_c$, the situation only becomes worse and all predictive power is
lost.

This is where models once again become important. There was a considerable
amount of model-dependent lore about the physics of heavy quarks before the
invention of heavy quark symmetry. Some of this lore, naturally, led to
heavy quark symmetry. Much of the remaining information can be used to tell
us more about things that HQET does not predict. However, HQET may show us
that some of the old model-dependent results were wrong.

As a concrete example, consider the Isgur-Wise function. HQET predicts the
normalization of the Isgur-Wise function at the zero-recoil point. Since
HQET does not predict the shape of the Isgur-Wise function, however, it
makes sense to calculate the Isgur-Wise function using a model. Heavy quark
symmetry predicts the normalization of the Isgur-Wise function, so a good
model must give the same result in the appropriate limit. In doing so, we
can gain insight into HQET, i.e., find a reasonable candidate Isgur-Wise
function. We can also gain insight into the model, i.e., see if the model
properly obeys the normalization condition on the Isgur-Wise function in
the appropriate limit.

In this paper I will focus on the functions that occur at subleading order
in $\lambar/m_Q$ in HQET, not on the Isgur-Wise function. These functions
are interesting for several reasons. One is that, theoretically, the
leading order results of heavy quark symmetry are somewhat trivial---in the
symmetry limit, the wave functions for $B^{(*)}$ and $D^{(*)}$ are simply
the same. The subleading effects are where the interesting effects of QCD
arise. More practically, it is important to understand the size of the
corrections to the symmetry limit in order to compare with experiment.
Ultimately, it is important to estimate the size of the correction to the
prediction used to extract $V_{cb}$. The relevant correction is actually
sub-subleading, however, so it is beyond the scope of this work.
Nonetheless, since the subleading effects need to be understood before
tackling the sub-subleading effects, this work is a step in the direction
of understanding the corrections relevant to extracting $V_{cb}$.

To calculate the subleading effects I have used a nonrelativistic quark
model. Specifically, I have used a slightly modified version of the
Isgur-Scora-Grinstein-Wise (ISGW) model \cite{ISGW}. There are several
reasons to choose this model. First, it is important start with a model
that is well established. It does not make sense to try to understand what
the old lore tells us about HQET by inventing an entirely {\em new} model
with which to study it. Secondly, the model is very simple. Since a major
goal of this work is to gain intuition about subleading effects in HQET,
simple results are very desirable. By performing the calculations using
approximate variational techniques, as was done by ISGW, all results can be
calculated in closed form. I consider this fundamentally more intuitive
than numerical results. One might argue that these symbolic results are
less accurate than involved numerical calculations. While this is true, it
is not necessarily useful to obtain very precise results based on less
precise assumptions. The ISGW model is, after all, a non-relativistic
model. Finally, because the ISGW model has been widely used, even to model
backgrounds in experiments, it is important to understand the ISGW model
itself. HQET may provide some insight as to the validity of its
assumptions.

The subleading mass effects in HQET have also been calculated using QCD sum
rules \cite{BagBalletal,NeubertSLSR,NeubertXi2,NeubertXi3} and a
Bethe-Salpeter-inspired relativistic quark model \cite{HoldSuth}. Dib and
Vera \cite{DibVera} have used the ISGW model to calculate subleading heavy
quark effects in $B\to\rho l\nu$ and $D\to\rho l\nu$ decays. Although such
heavy-light transitions are different from the heavy-heavy transitions
considered here, some of this work overlaps with theirs. A general
discussion of heavy quark symmetry-breaking effects using the
nonrelativistic quark model can be found in Ref.~\cite{Isgur}.

The rest of this paper proceeds as follows: \secref{sec:HQET} briefly
summarizes the results of HQET for heavy meson transitions to subleading
order in $\lambar/m_Q$, while \secref{sec:ISGW} summarizes the ISGW model
and shows that it is somewhat inconsistent with the results of HQET at
leading order. \secref{sec:ImpModel} shows how a small modification fixes
the discrepancy between ISGW and HQET at leading order, then shows how the
improved model can be used to calculate the effects of HQET at subleading
order with a minimum number of model restrictions. The following section
applies the assumptions and techniques of the ISGW model to obtain specific
results for the leading and subleading effects. Finally,
\secref{sec:discussion} compares these specific results with other
estimates and discusses their experimental testability and implications for
the ISGW model.

\section{Heavy Quark Effective Theory} \label{sec:HQET}

We start by defining the form factors for transitions between a heavy
pseudoscalar meson $B$ with four-velocity $v$ and a heavy pseudoscalar
(vector) meson $D$ ($D^*$) with four-velocity $v'$ and polarization vector
$\varepsilon$:
	\bes{eq:xidefns}
	{\bra{D(v')}V_\mu\ket{B(v)}\over\sqrt{m_Bm_D}}&=&
		\xi_+(w)(v+v')_\mu + \xi_-(w)(v-v')_\mu,
		\nn\\
	{\bra{D^*(v',\varepsilon)}V_\mu\ket{B(v)}\over\sqrt{m_Bm_{D^*}}}&=&
		i\xi_V(w)\epsilon_{\mu\nu\alpha\beta} {\varepsilon^*}^\nu v'^{\alpha}
		v^{\beta},\nn\\
	{\bra{D^*(v',\varepsilon)}A_\mu\ket{B(v)}\over\sqrt{m_Bm_{D^*}}}&=&
		\xi_{A_1}(w)(w+1)\varepsilon^*_\mu\nn\\
		&& - \xi_{A_2}(w)\varepsilon^*\cdot v v_\mu -
		\xi_{A_3}(w)\varepsilon^*\cdot vv'_\mu.
	\ees
The invariant momentum transfer is characterized by $w = v\cdot v'$. This
is related to the more traditional $q^2$ by $w = (m_B^2 + m_{D^{(*)}}^2 -
q^2)/(2m_B m_{D^{(*)}})$. The kinematic range for $B\to D^{(*)}$
semileptonic decay is $1\leq w \leq w_\sw{max}$, where $w_\sw{max} \approx
1.50~(1.59)$ for $D$ ($D^*$).

The results of HQET can be easily expressed using the trace formalism
\cite{TraceFormalism}. At leading order, the expression for the matrix
elements using this formalism is
	\be{eq:loTrace}
	{\bra{D^{(*)}(v')}\Gamma_\mu\ket{B(v)}\over\sqrt{m_Bm_{D^{(*)}}}}=
	-\xi(w)\tr[\overline{\bold{D}^{(*)}}\Gamma\bold{B}],
	\ee
where
	\be{eq:Ddef}
	\bold{D} = - {1+\slashchar{v}' \over 2}\gamma_5,
	\ee
	\be{eq:Dstardef}
	\bold{D^{*}} = {1+\slashchar{v}' \over 2}\slashchar{\varepsilon}
	\ee
and
	\be{eq:bardef}
	\overline{\bold{D}} = \gamma_0\bold{D}^{\dag}\gamma_0.
	\ee
$\bold{B}$ is defined similarly. Taking the traces and comparing with
\eqn{eq:xidefns} yields
	\be{eq:loOne}
	\xi_+ = \xi_V = \xi_{A_1} = \xi_{A_3} \equiv \xi
	\ee
and
	\be{eq:loTwo}
	\xi_- = \xi_{A_2} \equiv 0,
	\ee
where $\xi(w)$ is the Isgur-Wise function. This function is not calculable
in HQET or in perturbative QCD. It does, however, obey the important
normalization condition
	\be{eq:IWnorm}
	\xi(w=1) = 1.
	\ee

The leading corrections to \eqn{eq:loTrace} arise from radiative
corrections and the finite masses of the $b$ and $c$ quarks. The radiative
corrections are calculable in QCD, so it is not necessary to discuss them
from the standpoint of models. Therefore, all the calculations in this
paper are at tree level. It should be understood that the radiative
corrections should be added to the results obtained here. The leading
effect of the finite masses of the quarks is ${\cal
O}(\Lambda_\sw{QCD}/m_c)$. Since the ${\cal O}(\Lambda_\sw{QCD}/m_b)$
effects involve the same non-perturbative functions as the ${\cal
O}(\Lambda_\sw{QCD}/m_c)$ effects, I will only consider the ${\cal
O}(\Lambda_\sw{QCD}/m_c)$ effects here.

The relation between the weak current in the full theory and the weak
current in the effective theory has an expansion in powers of $1/m_Q$. At
${\cal O}(\Lambda_\sw{QCD}/m_c)$, the correction to the weak current in
the infinite-mass limit adds
	\bes{eq:slcurrent}
		&&-{\lambar\over4m_c}\tr\left\{\vphantom{\overline{\bold{D}^{(*)}}}
		\left[(v+v')^\alpha\psi_+(w) -
		(v-v')^\alpha\xi(w) \right. \right.\nn \\
	&&~~~+\left.\left.\left((1+w)\psi_+(w) +
		(1-w)\xi(w)\right)\gamma^\alpha\right]
		\overline{\bold{D}^{(*)}}\gamma_\alpha\Gamma\bold{B}\right\}
	\ees
to the right hand side of \eqn{eq:loTrace}. At this order, the constant
$\lambar$ is given by $\lambar = m_B - m_b = m_D - m_c$. The pseudoscalar
and vector meson masses are equal at this order, so ``$m_D$'' should be
taken as the spin averaged mass, $(3m_{D^*} + m_D)/4$, and similarly for
$m_B$. A new undetermined function, $\psi_+(w)$, also appears at this order
in HQET. This function is related to the function $\xi_+$\footnote{This
is not the same as the $\xi_+$ defined in \eqn{eq:xidefns}.} defined in
Ref.~\cite{LukeThm} by
	\be{null2} \xi^{\mbox{\scriptsize Luke}}_+ = {\lambar \over 2}\psi_+.
	\ee
It is also common in the literature to see the function $\xi_3$,
defined by
	\be{null3}
	\xi_3 = -{\lambar \over 2}[(1+w)\psi_+ + (1-w)\xi],
	\ee
instead of $\psi_+$.

At this order, modifications to the wave function appear
through insertions of the magnetic moment operator. The calculation
\cite{LukeThm} adds
the following traces to the right hand side of \eqn{eq:loTrace}:
	\bes{eq:wftrace}
	&&-{\lambar\over2m_c}\left\{
		\psi_1(w)\tr[\overline{\bold{D^{(*)}}}\Gamma\bold{B}] +
	i\psi_2(w)\tr[v_\mu\gamma_\nu
		\overline{\bold{D^{(*)}}}\sigma^{\mu\nu}
		\mbox{$1+\slashchar{v}'\over
		2$}\Gamma\bold{B}]
		\right. \nn \\
	&& + \left.\psi_3(w)\tr[\sigma_{\mu\nu}
		\overline{\bold{D^{(*)}}}\sigma^{\mu\nu}
		\mbox{$1+\slashchar{v}'\over
		2$}\Gamma\bold{B}] \right\}.
	\ees
The functions $\psi_{1,2,3}(w)$ are all uncalculable in perturbative QCD.
They are related to the functions $\chi_{1,2,3}(w)$ defined in
Ref.~\cite{LukeThm} by
	\be{eq:psi-chi}
	\chi_{1,2,3} = {\lambar \over 2}\psi_{1,2,3}.
	\ee
Luke \cite{LukeThm} has shown that $\psi_1(w)$ and $\psi_3(w)$ obey the
conditions
	\be{eq:lukethm}
	\psi_1(w=1) = \psi_3(w=1) = 0.
	\ee

Including both leading and subleading effects gives the following form
factors \cite{LukeThm,CorrectTraces}:
	\be{eq:xi+}
	\xi_+ = \xi + {\lambar\over2m_c}
		\left[\psi_1-2(w-1)\psi_2+6\psi_3\right],
	\ee
	\be{eq:xi-}
	\xi_- = {\lambar\over2m_c}\left[(w-2)\xi-(w+1)\psi_+\right],
	\ee
	\be{eq:xiV}
	\xi_V = \xi +
	{\lambar\over2m_c}\left[\xi+\psi_1-2\psi_3\right],
	\ee
	\be{eq:xiA1}
	\xi_{A_1} = \xi + {\lambar\over2m_c}\left[{w-1\over w+1}\xi+
	\psi_1-2\psi_3\right],
	\ee
	\be{eq:xiA2}
	\xi_{A_2} =
	{\lambar\over2m_c}\left[-\xi+2\psi_2+\psi_+ \right]
	\ee
and
	\be{eq:xiA3}
	\xi_{A_3} = \xi + {\lambar\over2m_c}
		\left[\psi_1 - 2\psi_2 - 2\psi_3 + \psi_+\right].
	\ee

\section{The ISGW Model} \label{sec:ISGW}

In the ISGW model \cite{ISGW}, an $S$-wave $Q\bar d$ meson $X$ is
represented by
	\be{eq:ISGWstate}
	{\ket{X(\vecp_X,s_X)}\over\sqrt{2m_X}} = \int d^3\vecp\,\sum_{s,\bar s}
		\phi_X(\vecp)\chi_{s\bar s}\ket{Q(\vecL_Q(\vecp_X,\vecp),s)
		\bar d(\vecL_d(\vecp_X,\vecp),\bar s)},
	\ee
where $\phi_X(\vecp)$ is the momentum wave function, $\chi_{s\bar s}$
couples the spins of the quarks into either a pseudoscalar meson or a
vector meson with polarization vector $\epsilon$, and $\vecL_{Q(d)}$ is
the momentum of the $Q$ ($d$) quark. In the non-relativistic
approximation, $\chi_{s\bar s}$ is a constant and the $\vecL$'s are given
by
	\be{eq:LQ}
	\vecL_Q(\vecp_X,\vecp) = \frac{m_Q}{m_X}\vecp_X + \vecp
	\ee
and
	\be{eq:Ld}
	\vecL_d(\vecp_X,\vecp) = \frac{m_d}{m_X}\vecp_X - \vecp.
	\ee
The vector and axial currents are the nonrelativistic limit of the
corresponding spinor bilinears.

The wave function $\phi_X(\vecp)$ comes from solving the Schr\"odinger
equation with a Hamiltonian of the form
	\be{eq:HISGW}
	{\cal H}_\sw{ISGW} = -{\nabla^2\over 2m_Q}  -{\nabla^2\over 2m_d}-
		\frac{4\alpha_s}{3r} + br+ c
	\ee
with $\alpha_s = 0.5$, $b = 0.18 \mbox{ GeV}^2$ and $c = -0.84 \mbox{
GeV}$. Instead of solving the Schr\"odinger equation numerically, ISGW
\cite{ISGW} use variational solutions with the harmonic oscillator wave
functions
	\be{eq:HO1S}
	\psi^{1S} = {\beta^{3/2}\over \pi^{3/4}}\exp\left(-{\beta^2 r^2 \over
		2}\right),
	\ee
	\be{eq:HO2S}
	\psi^{2S} = \sqrt{3\over2}{\beta^{3/2}\over \pi^{3/4}}
		\left(1-{2\over3}\beta^2r^2\right)\exp\left(-{\beta^2 r^2 \over
		2}\right),
	\ee
etc., as the basis. This allows the calculations to be done entirely
analytically instead of resorting to numerical calculations. This
approximation was checked in Ref.~\cite{ISGW} by showing that the mixing of
higher terms was small. It should be noted that, while this approximation
is probably quite reasonable for transitions and energy levels, it is
probably very poor for decay constants. This is because the variational
calculation provides a good fit {\em on average}, but it can be
significantly wrong at a given point. For example, since decay constants
depend only on the wave function at the origin, a variational calculation
of decay constants would not be reliable.

Calculating form factors using the ISGW model in the heavy quark limit,
i.e., $m_b = m_B$, $m_c = m_D$, $m_d \ll m_c,m_b$ and $\phi_B = \phi_D$,
yields
	\be{eq:ISGWnon-zero}
	\xi_+ = \xi_V = \xi_{A_3} \equiv F,
	\ee
	\be{eq:ISGWzero}
	\xi_- = \xi_{A_2} \equiv 0,
	\ee
and
	\be{eq:ISGWA1}
	\xi_{A_1} = \frac{2}{w+1}F.
	\ee
The function $F$ is given by
	\be{eq:F}
	F(w) = \exp\left[-{m_d^2\over 2 \beta^2}(w-1)\right],
	\ee
where $\beta$ is the width of the meson wave function given in
\eqn{eq:HO1S}. Comparing Eqs.~(\ref{eq:ISGWnon-zero})--(\ref{eq:ISGWA1})
with Eqs.~(\ref{eq:loOne}) and (\ref{eq:loTwo}), we see that the ISGW model
agrees with the heavy quark result at zero recoil ($w=1$), but the form
factors do not obey the symmetry for $w\neq1$. This failure is easy to
understand---in the rest frame of the $B$ meson, $w=E_D/m_D$. A completely
non-relativistic approximation therefore cannot be expected to produce
proper $w$-dependences. We shall see in the next section that this
particular problem is easily overcome.

\section{The Improved Model} \label{sec:ImpModel}

Before attempting to calculate the subleading effects in HQET with the
nonrelativistic quark model, it is necessary to find a way to make the
nonrelativistic quark model agree exactly with the predictions of heavy
quark symmetry at leading order. For example, it, would not otherwise be
clear {\em a priori} whether we should identify the Isgur-Wise function
with $F$ or $2F/(w+1)$. Fortunately, the fix is simple. In the improved
model, I define the $Q\bar d$ pseudoscalar (vector) mesons $H$ ($H^*$) to
be
	\be{eq:PseudoDef}
	\ket{H(v)} = -\int d^3\vecp \sum_{r,s} \phi_H(\vecp)
		\bar u_Q(\vecL_Q(v,\vecp),s)\gamma^5
		v_d(\vecL_{\bar d}(v,\vecp),r)
	\ee
and
	\be{eq:VectorDef}
	\ket{H^*(v,\varepsilon)} = \int d^3\vecp \sum_{r,s} \phi_H(\vecp)
		\bar u_Q(\vecL_Q(v,\vecp),s)\slashchar{\varepsilon}
		v_d(\vecL_{\bar d}(v,\vecp),r).
	\ee
The corresponding transition current is given by
	\be{eq:Cur}
	J^\mu_{b\to c}=\bar u_c\Gamma^\mu u_b,
	\ee
where $\Gamma_\mu = (\gamma_\mu, \gamma^5\gamma_\mu)$ for $J_\mu = (V_\mu,
A_\mu)$. In calculating transition matrix elements, one must also insert
momentum-conserving delta functions for the quarks. By writing the quarks
in terms of spinors and gamma matrices, it is easier to keep track of the
Lorentz properties of the heavy quark. The authors of Ref.~\cite{ISGW}
concluded that the inclusion of these effects was not necessary. Their
inclusion is necessary, however, to get the spin symmetry in the heavy
quark limit. Since, in the infinite mass limit, the momentum of the heavy
quark is just the momentum of the meson, keeping the exact transformation
properties of the heavy quark is very simple in the heavy quark limit. This
should become obvious when we do an explicit calculation in this limit.

Throughout this section, I will continue to write expressions in the most
general terms, i.e., without making any unnecessarily specific assumptions
about the form of $\phi_H$, the non-relativistic form of $\vecL$, etc. The
following section applies the assumptions of the ISGW model to the more
general results of this section.

The heavy quark limit of this model is straightforward. The heavy quark
limit of the wave function, $\phi_H\stackrel{m_Q \to
\infty}{\longrightarrow}\phi_\infty$, is a solution to a wave equation with
a Hamiltonian of the form
	\be{eq:IMH}
	{\cal H}_\sw{HQ} = T_d + V_\sw{spinless},
	\ee
where $T_d$ is the kinetic energy of the $d$ quark and $V_\sw{spinless}$ is
a quark-antiquark binding potential that does not depend on the spins of
the quarks. As stated before, the momentum of the heavy quark becomes
simply the momentum of the meson. We can now do a sample calculation:
	\bes{eq:samp}
	\lefteqn{\bra{D(v')}V^\mu \ket{B(v)} = \int d^3\vecp' d^3\vecp\,
		\phi_\infty(\vecp') \phi_\infty(\vecp) \sum_{rr'ss'}}
		\delta^3(\vecL_d-\vecL_d')\\
		&&\bar v_d(\vecL_d',r')
		\gamma^5 u_c(\vecL_c,s')\bar u_c(\vecL_c,s')\gamma^\mu
		u_b(\vecL_b,s)\bar u_b(\vecL_b,s)\gamma^5 v_d(\vecL_d,r)\nn\\
		&=&\tr\left[\gamma^5\frac{(1+\slashchar{v}')}{2}\gamma^\mu
		\frac{(1+\slashchar{v})}{2}\gamma^5\right]
		\int d^3\vecp' d^3\vecp\,\phi_\infty(\vecp')\phi_\infty(\vecp)
		\delta^3(\vecL_d-\vecL_d')\nn
	\ees
The calculation with a $D^*$ in the final state just involves substituting
$-\slashchar{\varepsilon}$ for the leftmost $\gamma^5$ in the above
expression and inserting the appropriate current. It is clear that we have
reproduced the formalism of \eqn{eq:loTrace} with an Isgur-Wise function
given by
	\be{eq:IMIW}
	\xi(w) = \int d^3\vecp' d^3\vecp\,\phi_\infty(\vecp')\phi_\infty(\vecp)
		\delta^3(\vecL_d-\vecL_d').
	\ee
The normalization of the wave function $\phi_\infty$ gives the required
normalization of the Isgur-Wise function. The improved model thus exactly
and transparently reproduces the results of HQET, while allowing us to
calculate the Isgur-Wise function.

To go beyond leading order, start with the current correction. To calculate
the current correction in the quark model, we need to find the relation
between the spinor in the heavy quark limit, $u_c(m_cv')$, and the full
spinor, $u_c(p_c)$ to order $1/m_c$. Following Luke \cite{LukeThm}, we
write
	\be{eq:kdef}
	p_c^\mu = m_c v'^\mu + k^\mu
	\ee
and use the equation of motion
	\be{eq:eom}
	(\slashchar{p}_c - m_c)u_c = 0
	\ee
to write
	\bes{eq:slu}
	u_c(p_c)&=&\left({1+\slashchar{v}'\over 2}\right)u_c(p_c) +
		\left(1-\slashchar{v}'\over 2\right)u_c(p_c) \nn \\
	u_c(p_c)&=&\left({1 + \slashchar{v}' + \slashchar{k}/m_c\over2}\right)
		\left({1+\slashchar{v}'\over2}\right)u_c(p_c) +
		\left({1 + \slashchar{v}' + \slashchar{k}/m_c\over2}\right)
		\left({1-\slashchar{v}'\over2}\right)u_c(p_c) \nn \\
	&=&\left(1 + {\slashchar{k}\over2m_c}\right)
		\left({1+\slashchar{v}'\over2}\right)u_c(m_cv') +
		{\cal O}(k^2/m_c^2).
	\ees
The current given in \eqn{eq:Cur} is now
	\be{eq:CurCorr}
	J^\mu_{b\to c}=\bar u_c\left({1+\slashchar{v}'\over 2}\right)
		\left(1 + {\slashchar{k}\over2m_c}\right)\Gamma^\mu u_b.
	\ee
When we insert this current into the trace in \eqn{eq:samp}, the
$\slashchar{k}/2m_c$ piece will give a contribution of the form
	\be{eq:CurQM}
	-\tr[l^\alpha\overline{\bold{D}^{(*)}}\gamma_\alpha
		\Gamma\bold{B}],
	\ee
where
	\be{eq:lQM}
	\bold{l} = {1\over2m_c}\int d^3\vecp' d^3\vecp\,
		\phi_\infty^*(\vecp')
		\phi_\infty(\vecp)\bold{k}(\vecp')\delta^3(\vecL_d-\vecL_d'),
	\ee
where $\bold{k}(\vecp')$ is the residual momentum of the $c$-quark given
by \eqn{eq:kdef}. Comparing \eqn{eq:CurQM} with the HQET result in
\eqn{eq:slcurrent} yields
	\be{eq:lform}
	l^\mu = {\lambar\over4m_c}\left\{(v+v')^\mu\psi_+ - (v-v')^\mu\xi
		+ \left[(1+w)\psi_+ + (1-w)\xi\right]\gamma^\mu\right\}.
	\ee
Therefore we can calculate $\psi_+$ by performing the integral in
\eqn{eq:lQM} and comparing the result with \eqn{eq:lform}. This method also
yields a consistency check, because $l$ must have the form given in
\eqn{eq:lform}.

A complete calculation to ${\cal O}(1/m_c)$ must include corrections to
the heavy-quark limit of the wave function. In general, these corrections
will be of the form
	\be{eq:wavefnform}
	\phi_D = \phi_\infty + \phi^1,
	\ee
where $\phi_1$ is the effect of the first-order perturbation. When this is
inserted in the calculation in \eqn{eq:samp}, it will add a contribution
	\be{eq:extra}
	\int d^3\vecp' d^3\vecp\,\phi_\infty(\vecp')\phi^1(\vecp)
		\delta^3(\vecL_d-\vecL_d')
	\ee
to the integral that formerly represented only the Isgur-Wise function.

Calculating these corrections in perturbation theory is straightforward.
The full Hamiltonian has a kinetic energy term for the heavy quark
	\be{eq:ke}
	{\cal H}_\sw{KE} = -{\nabla^2 \over 2m_Q}
	\ee
Of course, we could use the full relativistic expression for the heavy
quark kinetic energy, but the difference between the nonrelativistic
kinetic energy and the relativistic kinetic energy is higher order in
$1/m_Q$, so we are completely justified in using the nonrelativistic
expression in this context.

Denoting the $n^\sw{\em th}$ wave function of the complete set of basis
states of the solution to the unperturbed Hamiltonian by $\phi_n$ and
applying ordinary time-independent perturbation theory, we have
	\be{eq:KEeqn}
	\phi_D = \phi_\infty + \phi_\sw{KE}^1,
	\ee
where
	\be{eq:phi1KEdef}
	\phi_\sw{KE}^1(\vecr) = \sum_{n\neq\infty} {\phi_n(\vecr)\over E_n -
		E_\infty}\int
		d^3\vecr' \phi_n^*(\vecr') {\cal H}_\sw{KE} \phi_\infty(\vecr').
	\ee

The Hamiltonian also has a spin-spin interaction term
	\be{eq:HSS}
	{\cal H}_\sw{SS} = {\bold{S}_d\cdot\bold{S}_Q\over
		m_d m_Q} V_\sw{SS}(r),
	\ee
where $\bold{S} = \bold{\sigma}/2$ is the spin of the quark. Once again
applying ordinary time-independent perturbation theory, we now have
	\be{eq:xxx3}
	\phi_D = \phi_\infty + \phi_\sw{KE}^1+
		\langle\bold{S}_c\cdot\bold{S}_d\rangle\phi_\sw{SS}^1,
	\ee
where
	\be{eq:phi1SSdef}
	\phi_\sw{SS}^1(\vecr)=
		\sum_{n\neq\infty} {\phi_n(\vecr)\over E_n - E_\infty}\int
		d^3\vecr' \phi_n^*(\vecr') {V_\sw{SS}(r)\over m_d m_c}
		\phi_\infty(\vecr'),
	\ee
with $\langle\bold{S}_c\cdot\bold{S}_d\rangle =
(-\frac{3}{4},\frac{1}{4})$ for $(D,D^*)$.

General QCD-inspired Hamiltonians for the quark-quark interaction also
contain spin-orbit and tensor interactions. These terms have no effect
when applied as perturbations to $S$-wave states, so they do not
contribute to this calculation.

The first term in \eqn{eq:wftrace}, which involves $\psi_1$, breaks the
flavor symmetry, but leaves the spin symmetry unchanged. The effects of
\eqn{eq:ke} in the quark model calculation are precisely the same.
In the quark model calculation, the effect of \eqn{eq:KEeqn} is
	\be{eq:qmpsi1effect}
	\xi \to \xi +
		\int d^3\vecp' d^3\vecp\,\phi_\sw{KE}^1(\vecp')\phi_\infty(\vecp)
		\delta^3(\vecL_d-\vecL_d')
	\ee
whereas in HQET,
	\be{eq:HQETqmpis1effect}
	\xi \to \xi +
		{\lambar \over 2m_c}\psi_1.
	\ee
We can therefore identify
	\be{eq:psi1eqn}
	\psi_1 ={2m_c\over \lambar}
		\int d^3\vecp' d^3\vecp\,\phi_\sw{KE}^1(\vecp')\phi_\infty(\vecp)
		\delta^3(\vecL_d-\vecL_d').
	\ee

The third symmetry-breaking term in \eqn{eq:wftrace}, which involves
$\psi_3$, has the spin dependence of the hyperfine interaction in
\eqn{eq:HSS}. This can be seen by explicit calculation, which shows that
both forms have the net effect of multiplying the original trace structure
by a factor proportional to $+3~(-1)$ for pseudoscalar (vector) mesons.
Specifically, the quark model calculation gives
	\be{eq:qmpsi3effect}
	\xi \to \xi + \left(
		\begin{array}{c}
		-3/4\mbox{ for $D$}  \\
		+1/4\mbox{ for $D^*$}
		\end{array} \right)
		\int d^3\vecp' d^3\vecp\,\phi_\sw{SS}^1(\vecp')\phi_\infty(\vecp)
		\delta^3(\vecL_d-\vecL_d'),
	\ee
whereas the HQET result is
	\be{eq:HQETpsi3effect}
	\xi \to \xi+\left(
		\begin{array}{c}
		+6\\
		-2
		\end{array} \right)
		{\lambar \over 2m_c}\psi_3.
	\ee

This implies
	\be{eq:psi3eqn}
	\psi_3 = {-m_Q\over 4\lambar}
		\int d^3\vecp' d^3\vecp\,\phi_\sw{SS}^1(\vecp')\phi_\infty(\vecp)
		\delta^3(\vecL_d-\vecL_d').
	\ee

The functions $\psi_1$ and $\psi_3$ are constrained by \eqn{eq:lukethm}. It
is interesting to note that this constraint is automatically
satisfied---not because of the details of the model, but because of the
general properties of time-independent perturbation theory. In
general\footnote{This proof follows that given in Ref.~\cite{CoTan}.}, we
want to find the eigenstates $\ket{A(\lambda)}$ of a Hamiltonian
$H(\lambda)$, where $\lambda$ is an arbitrarily small real parameter. We
expand $\ket{A(\lambda)}$ as follows:
	\be{eq:nrpt}
	\ket{A(\lambda)} = \ket{0} + \lambda\ket{1} + {\cal O}(\lambda^2).
	\ee
$\ket{A(\lambda)}$ is only determined up to an arbitrary complex constant.
We can fix the magnitude and the phase of the constant by setting $\langle
A(\lambda) | A(\lambda) \rangle = 1$ and demanding that $\langle A(\lambda)
| 0 \rangle$ be real. Then
	\bes{Anorm}
	\langle A(\lambda) | A(\lambda) \rangle&=&\langle 0 | 0 \rangle +
		\lambda (\langle 0|1 \rangle + \langle 1|0 \rangle) + {\cal
		O}(\lambda^2) \nn \\
	&=&\langle 0 | 0 \rangle +
		2\lambda \mbox{Re}\langle 0|1 \rangle + {\cal O}(\lambda^2).
	\ees
Since $\lambda$ is an arbitrary parameter, the normalization of
$\ket{A(\lambda)}$ and the phase convention requires
	\be{eq:noOverlap}
	\langle 0|1\rangle = 0.
	\ee
At zero recoil, the integrals in Eqs.~(\ref{eq:psi1eqn}) and
(\ref{eq:psi3eqn}) just correspond to the overlap of the first-order
correction to the wave function ($\ket{1}$) with the leading-order wave
function ($\ket{0}$). The vanishing of $\psi_1$ and $\psi_3$ at zero-recoil
is just a special case of the general result in \eqn{eq:noOverlap}.

After applying the corrections to the current and the wave function, all of
the new unknown functions that appear at ${\cal O}(1/m_Q)$ are accounted
for, with the sole exception of $\psi_2$. The model not does provide for
any effects of the particular symmetry breaking form associated with
$\psi_2$, so $\psi_2(w)\equiv0$ in this model.

\section{Specific Results}

This section applies the nonrelativistic formalism of the ISGW model to
results of the previous section to obtain numerical predictions for the
form factors.

We must start by calculating the wave function $\phi_\infty$ using the
Hamiltonian
	\be{eq:HHQ}
	{\cal H}_\sw{HQ} = -{\nabla^2\over 2m_d}-
		\frac{4\alpha_s}{3r} + br+ c.
	\ee
The variational calculation using the trial wave function in \eqn{eq:HO1S}
gives $\beta = 420\mbox{ MeV}$ using the ISGW parameters. (In addition to
those parameters listed in \secref{sec:ISGW}, ISGW use $m_d = 330\mbox{
MeV}$.) Dib and Vera \cite{DibVera} have obtained the same value of $\beta$
in a similar calculation.

Using \eqn{eq:IMIW}, we now obtain the Isgur-Wise function:
	\be{eq:IWresult}
	\xi(w) = \exp\left[-{m_d^2\over 2 \beta^2}(w-1)\right].
	\ee
In the standard parameterization $\xi(w) = 1 - \rho^2(w-1) + {\cal
O}[(w-1)^2]$, this corresponds to $\rho^2 = m_d^2/2\beta^2$. Unfortunately,
as was discussed in Ref.~\cite{ISGW}, the resulting shape is unreliable.
The problem is that, in the rest frame of the $B$, the integral in
\eqn{eq:IMIW} amounts to unity plus a correction of ${\cal
O}(\bold{p}_D^2/m_D^2)$, which we cannot expect to calculate reliably in a
nonrelativistic model.

The authors of Ref.~\cite{ISGW} compensate for this deficiency by
introducing a phenomenological parameter $\kappa$, resulting in the
expression in \eqn{eq:F}. They obtained $\kappa=0.7$ from a fit to the pion
form factor. Since we are considering this model from a heavy quark
standpoint, it is not clear what fitting to the pion form factor even means.
Instead, we can use a fit from Ref.~\cite{mySemiLep}, which obtained
$\rho = 0.93\pm0.10$ using the exponential form for the Isgur-Wise
function. This gives $\kappa = 0.60\pm0.06$, which is remarkably close to
the value obtained from the pion form factor. For the rest of the paper I
use
	\be{eq:phenIW}
	\xi(w) = \exp\left[-{m_d^2\over 2\kappa^2 \beta^2}(w-1)\right]
	\ee
with $\kappa = 0.6$. Of course, it would be better to be able to {\em
predict} the shape of the Isgur-Wise function, but that is not the focus of
this work. The problem of accurately calculating the Isgur-Wise function in
the context of the nonrelativistic quark model has been considered by
others \cite{Close}.

To calculate the corrections to the current, we need to evaluate the
integral in \eqn{eq:lQM}. In the rest frame of the $B$ meson,
this reduces to
	\bes{eq:CurIntegralResult}
	\bold{l} &=& {1\over2m_c}\int d^3\vecp\,\phi_\infty^*
		(\vecp + m_d \bold{v}')
		\phi_\infty(\vecp)\vecp \nn \\
	&=& {m_d\over4m_c}\, \bold{v}'
		\exp\left[-{m_d^2\over 2 \beta^2}(w-1)\right].
	\ees
This is to be compared with \eqn{eq:lform}, which becomes
	\be{eq:lform2}
	\bold{l} = {\lambar\over4m_c} \left\{(\psi_+ + \xi)\bold{v}' +
		 \left[(1+w)\psi_+ + (1-w)\xi\right]\bold{\gamma}\right\}.
	\ee
Solving Eqs.~(\ref{eq:CurIntegralResult}) and (\ref{eq:lform2}) for the
unknowns $\lambar$ and $\psi_+$ gives
	\be{eq:lambarresult}
	\lambar = m_d
	\ee
and
	\be{eq:psi+result}
	\psi_+(w)\equiv0.
	\ee
Unfortunately, it also gives an inconsistency, because it seems that
\eqn{eq:CurIntegralResult} is missing the term $(1-w)\xi\bold{\gamma}$.
This term is suppressed by a factor proportional to $(1-w)$ relative to the
other terms, however, so we cannot expect to get it right in a
nonrelativistic calculation. This again demonstrates the limits of the
nonrelativistic model.

Next, we calculate the correction to the wave function $\phi^1_\sw{KE}$.
The second derivative in the kinetic energy brings down two powers of
$r$, so the effect of the perturbation is to mix in a piece
proportional to $\psi^{2S}$. Using \eqn{eq:phi1KEdef} gives
	\be{eq:phi1KEresult}
		\phi^1_\sw{KE}(\vecr) = -\sqrt{3\over2}{\beta^2
			\over 2m_c(E_{2S} - E_{1S})}\psi^{2S}(\vecr).
	\ee
If we knew the masses of the $D(2S)$ or $B(2S)$ mesons, we could use get
the quantity $E_{2S} - E_{1S}$ directly from experiment. Since these
numbers are not available, I use the ISGW model calculation, which gives
	\be{eq:deltaE}
	E_{2S} - E_{1S} = {\beta^2 \over m_d} + {b \over \beta \sqrt{\pi}} +
		{4 \alpha_s \beta \over 9 \sqrt{\pi}}.
	\ee
Numerically, this is $829\mbox{ MeV}$. Eichten, Hill and Quigg
\cite{Quigg} have recently calculated the $2S$--$1S$ splittings in $D$ and
$B$ systems using the Buchm\"uller-Tye potential \cite{BuchTye}. They
obtain approximately $740\mbox{ MeV}$ and $720\mbox{ MeV}$ in the $D$ and
$B$ systems, respectively. This means that the $829\mbox{ MeV}$ result is
probably an overestimate, but the discrepancy is less than $15\%$---which
is acceptable considering the approximate nature of the model.

Plugging \eqn{eq:phi1KEresult} into \eqn{eq:psi1eqn} gives
	\be{eq:firstpsi1result}
	\psi_1(w) = {m_d \over 2 (E_{2S} - E_{1S})}(w-1)
		\exp\left[-{m_d^2\over 2 \beta^2}(w-1)\right].
	\ee
Given the earlier caveats about determining the shape of the form factors
with this model, I write this as
	\be{eq:psi1result}
	\psi_1(w) = {m_d \over 2 (E_{2S} - E_{1S})}(w-1)\xi(w)
	\ee
and use the form for $\xi(w)$ given in \eqn{eq:phenIW}. Numerically,
	\be{eq:numresult1}
	{m_d\over 2(E_{2S} - E_{1S})} \approx 0.20.
	\ee

Finally, we calculate the hyperfine correction to the wave function,
$\phi^1_\sw{SS}$. For the potential in \eqn{eq:HISGW}, the appropriate form
for $V_\sw{SS}(\bold{r})$ in \eqn{eq:HSS} is \cite{OlsOlssWill}
	\be{eq:Vssform}
	V_\sw{SS}(\bold{r}) = g \delta^3 (\bold{r}),
	\ee
where $g = 32\pi\alpha_s/9$. The $\delta$-function, when inserted in
\eqn{eq:phi1SSdef}, will mix in pieces of {\em all} $nS$ states, where
$n>1$. Fortunately, the $(n+1)S$ states give a contribution to the form
factor suppressed relative to the $nS$ state by a factor of $(w-1)$. Since
we miss such terms in the nonrelativistic approximation, it only makes
sense to include the dominant $2S$ contribution. Now
	\be{eq:firstphi1SSresult}
		\phi^1_\sw{SS}(\vecr) = {g
			\over m_dm_c (E_{2S} - E_{1S})}\psi^{2S*}(0)\psi^{1S}(0)
		\psi^{2S}(\vecr).
	\ee
Unfortunately, this calculation of $\phi^1_\sw{SS}$ is much less reliable
than the previous $\phi^1_\sw{KE}$ calculation because $\phi^1_\sw{SS}$
depends on the wave functions at a single point, instead of an average over
all of space. As was discussed in \secref{sec:ISGW}, the variational
calculation should not be expected to be very accurate for this type of
calculation.

Fortunately, we have experimental information that reduces the dependence
on knowledge of the wave function at the origin. The term ${\cal
H}_\sw{SS}$ gives rise to the $D^*$-$D$ mass splitting:
	\be{eq:getg}
	\Delta_D \equiv m_{D^*} - m_{D} = {g \over m_d m_c}|\phi(0)|^2.
	\ee
To use this result, we need to know the relation between $\psi^{2S}(0)$
and $\psi^{1S}(0)$. Using the harmonic oscillator wave functions gives
	\be{eq:hoOrigin}
	\psi^{2S}(0)= \sqrt{\frac{3}{2}}\psi^{1S}(0).
	\ee
This is almost certainly an overestimate, however.  An exact result
for potential models \cite{Martin} states that
	\be{eq:origin}
	{d^2 \over dr^2}V(r) \gtlt 0 \Rightarrow
		|\psi^{2S}(0)|^2 \gtlt |\psi^{1S}(0)|^2.
	\ee
The second derivative of the harmonic oscillator potential is a positive
constant, so \eqn{eq:hoOrigin} is in accord with the relation in
\eqn{eq:origin}. The second derivative of the potential from the
Hamiltonian in \eqn{eq:HISGW}, however, is always negative, so the exact
solution must have $\psi^{2S}(0)<\psi^{1S}(0)$. If we had a pure linear
potential, the $1S$ and $2S$ wave functions would be equal at the origin.
Since these heavy-light systems are dominated by the linear part of the
potential, I use
	\be{eq:linOrigin}
	\psi^{2S}(0)= \psi^{1S}(0)
	\ee
with the understanding that this must be somewhat of an overestimate.

We now have
	\be{eq:phi1SSresult}
		\phi^1_\sw{SS}(\vecr) = {\Delta_D
		\over E_{2S} - E_{1S}}\psi^{2S}(\vecr).
	\ee
Had we used \eqn{eq:hoOrigin}, the above formula and
Eqs.~(\ref{eq:phi1KEresult}) and (\ref{eq:deltaE}) would agree with similar
calculations performed by Dib and Vera \cite{DibVera}. As it stands, only
\eqn{eq:phi1SSresult} disagrees with their results.

Following the same procedure as we used for $\psi_1$ above,
	\be{eq:psi3result}
		\psi_3(w) = {m_cm_d\Delta_D\over 4\sqrt{6}\beta^2(E_{2S} - E_{1S})}
			(w-1)\xi(w).
	\ee
Once again, I use the form of $\xi(w)$ given in \eqn{eq:phenIW}.
Numerically,
	\be{eq:numresult2}
	{m_cm_d\Delta_D\over 4\sqrt{6}\beta^2(E_{2S} - E_{1S})} \approx
		0.053.
	\ee
Here I have used $m_c = 1.64\mbox{ GeV}$, instead of the value from
Ref.~\cite{ISGW} ($1.82\mbox{ GeV}$) for consistency with the relation
$\lambar = (3m_{D^*} + m_D)/4 - m_c$.

\section{Discussion} \label{sec:discussion}
\begin{figure}
	\epsfxsize=\hsize
	\epsffile{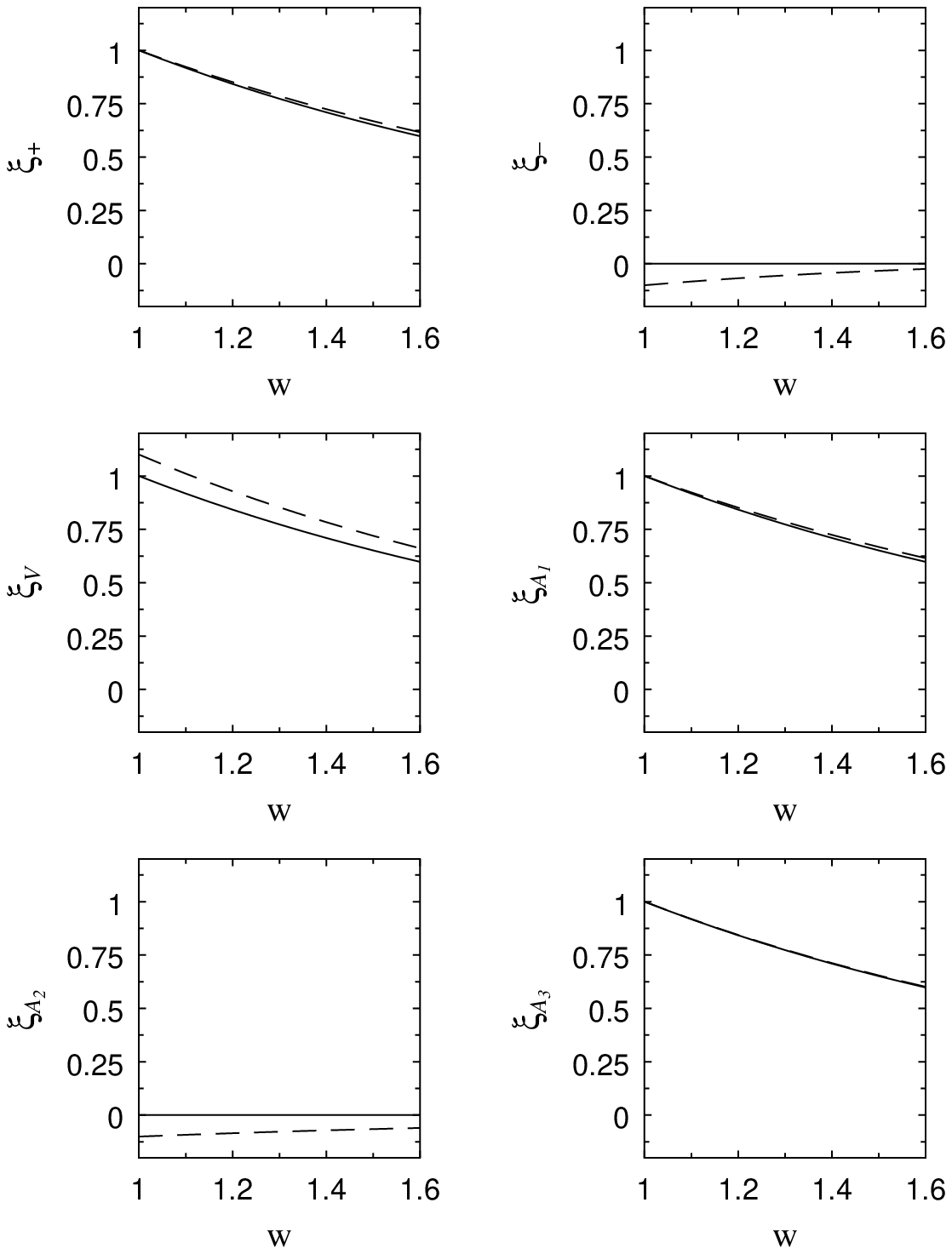}
	\caption{$B\to D^{(*)}$ form factors with $1/m_c$ corrections (dashed
	lines) and without (solid lines).}
	\label{fig:form-factors}
	\end{figure}

The net effect of these results on the $B\to D^{(*)}$ form factors is
shown in \fig{fig:form-factors}.
The results have several general features. The most important result is
probably the size of $\lambar$. With the result $\lambar = 330\mbox{ MeV}$,
the expansion parameter that multiplies the subleading effects is
$\lambar/(2m_c) \approx 10\%$. In general, we should expect that $\lambar =
m_d$ is determined to typical quark model accuracy, probably $20\%$. It
should be noted that the fact that the quark model implies $\lambar=m_d$
has been common knowledge, probably since the inception of heavy quark
symmetry, but it is reassuring to see the result come out of an explicit
calculation.

The functions $\psi_1$ and $\psi_3$ both are positive. While the
calculation of $\psi_3$ is somewhat unreliable, the sign and approximate
magnitude are well determined. In addition, many uncertainties cancel in
the ratio
	\bes{eq:psiratio}
	{\psi_1(w) \over \psi_3(w)}&=&{2\sqrt{6}\beta^2 \over m_c \Delta_D}
		{\psi^{2S}(0) \over \psi^{1S}(0)}\nn \\
		& \approx & 3.7 {\psi^{2S}(0) \over \psi^{1S}(0)},
	\ees
which, along with \eqn{eq:origin}, implies that $\psi_1$ should be {\em at
least} $3.7$ times as large as $\psi_3$.

The largest corrections to the symmetry limit are those from the current
correction which are proportional to the Isgur-Wise function. In the plot
the effects of $\psi_1$ and $\psi_3$ are barely noticeable. They are
suppressed with respect to the current corrections partly because they
contain small leading constants and partly because they both carry a factor
of $(w-1)$, which never becomes greater than $0.6$ throughout the
kinematic range accessible in $B\to D^{(*)}$ semileptonic transitions.

In Ref.~\cite{mySemiLep}, we analyzed $D\to K^{(*)}$ semileptonic decays
using HQET, treating the strange quark as heavy. Although it is not clear
that this approach should work at all, it seemed reasonably successful. We
used the measured form factors for $D\to K^{(*)}$ to extract
phenomenological estimates of $\psi_+, \psi_1,\psi_2$ and $\psi_3$ at
various kinematic points. \tabref{tab:semiLep} compares these results with
the results of Ref.~\cite{mySemiLep}. Two cautions are in order. Treating
the strange quark as heavy simply may not be reliable; the error bars in
the table do not include systematic errors due to treating the strange
quark as heavy. Also, the results of the present work are not reliable at
$w=2$. The non-relativistic approximation relies on the quantity $(w-1)$
being small. Since this is obviously not true at $w=2$, the present results
should only be considered order of magnitude estimates at $w=2$. All of
these cautions aside, however, the estimates agree reasonably well. In
particular, the result that $\psi_2$ and $\psi_+$ vanish agrees with the
phenomenological result. The phenomenological result is also consistent
with the approximate magnitude of $\psi_1$ and $\psi_3$, in addition to the
result in \eqn{eq:psiratio}, albeit with large errors.

The errors on $\psi_1$ and $\psi_3$ can be reduced if we assume $\psi_+(w)
= \psi_2(w) = 0.$ Under this assumption, the result becomes
$\psi_1(2)+6\psi_3(2)=1.9\pm0.2~(0.62\pm0.06)$ for $\lambar=200\mbox{
MeV}~(400\mbox{ MeV})$. The calculated signs of $\psi_1$ and $\psi_3$ are
consistent with this result.

It should be noted that an important motivation for the analysis of
Ref.~\cite{mySemiLep} was the failure of the non-relativistic quark models
and heavy quark symmetry to properly predict the ratio $\Gamma(D\to
K^*l\nu)/\Gamma(D\to Kl\nu)$. It turns out that radiative QCD corrections
made a crucial contribution. The ISGW model fails to reproduce the ratio
mostly because it lacks these corrections. The heavy-mass effects due to
$\psi_1$ and $\psi_3$ also contribute, but to a lesser degree. The
agreement between the values extracted from experiment and the calculation
in \tabref{tab:semiLep}, particularly the agreement of the signs of
$\psi_1$ and $\psi_3$, indicates the effects calculated here work in the
proper direction to fix the discrepancy.

\begin{table}
\caption{Comparison of the present work with a phenomenological analysis
of $D \to K^{(*)}$ semileptonic decays.}
\protect\label{tab:semiLep}
\begin{center}
\begin{tabular}{cccc}
\hline\hline
&Ref.~\protect\cite{mySemiLep} with &
 Ref.~\protect\cite{mySemiLep} with &\\
&$\lambar = 200\mbox{ MeV}$ & $\lambar = 400\mbox{ MeV}$ & This Work  \\
\hline
$\psi_+(1)$ & $0.2\pm1.3$ & $-0.3\pm0.4$ & $0\ap\ad\ad$ \\
$\psi_1(2)$ & $0.9\pm0.5$ & $\am0.1\pm0.2$ & $0.08$ \\
$\psi_2(1)$ & $1.1\pm1.9$ & $-0.4\pm0.6$ & $0\ap\ad\ad$ \\
$\psi_3(2)$ & $0.5\pm0.3$ & $\am0.0\pm0.1$ & $0.02$ \\
\hline\hline
\end{tabular}
\end{center}
\end{table}

These calculations can also be compared with similar calculations done with
QCD sum rules. There are two sum rule calculations of $\lambar$. One
obtains $\lambar = 500 \pm100\mbox{ MeV}$ \cite{BagBalletal} while the
other obtains $\lambar = 570 \pm 70\mbox{ MeV}$ \cite{NeubertReview}. These
numbers are somewhat larger than the results obtained here.

\tabref{tab:SumRules} compares the sum rule calculations of
Refs.~\cite{NeubertSLSR,NeubertXi2,NeubertXi3} with the results of this
work. The sum rule results and the quark model results for $\psi_3$ agree.
Qualitatively, both models predict that $\psi_1$ is several times larger
than $\psi_3$ and that the two have the same sign. However, the sum rule
result for $\psi_1$ is substantially larger than the quark model result.
Since $\psi_1$ represents the effects of the kinetic energy of the heavy
quark, we can see that sum rules indicate considerably more motion of the
heavy quark inside the meson than the quark model does. The disagreement
with $\psi_+$ and $\psi_2$ seems more fundamental. However, the first
calculation of $\psi_2$, which used the standard sum rule assumptions
\cite{NeubertSLSR}, found the same result obtained here, i.e.,
$\psi_2(w)\equiv0$. It was only after the two-loop radiative corrections to
the quark loop, the one-loop radiative corrections to the quark condensate,
and the gluon condensate \cite{NeubertXi2} were included that a non-zero
result for $\psi_2$ was found. Although the exact connection between sum
rules and the nonrelativistic quark model is not clear, it is
understandable that the above terms do not appear in the nonrelativistic
quark model. Nonetheless, even the improved sum rule results for $\psi_2$
are quite small.

\begin{table}
\caption{Comparison of the present work with sum rule calculations from
Refs.~\protect\cite{NeubertSLSR}, \protect\cite{NeubertXi2} and
\protect\cite{NeubertXi3}.}
\protect\label{tab:SumRules}
\begin{center}
\begin{tabular}{ccc}
\hline\hline
&Sum Rules & This Work  \\
\hline
$\psi_+(1.0)$ \protect\cite{NeubertSLSR} & $\ad-0.33\pm0.05\ad$ &
	 $0\ap\ad\ad$ \\
$\psi_+(1.6)$ \protect\cite{NeubertSLSR} & $\ad\am0.00\pm0.04\ad$ &
	 $0\ap\ad\ad$ \\
$\psi_1(1.6)$ \protect\cite{NeubertSLSR} & $\ad\ad\am1.0\pm0.3\ad\ad$ &
	 $0.07$ \\
$\psi_2(1.0)$ \protect\cite{NeubertSLSR} & $\ad-0.08\pm0.02\ad$  &
	 $0\ap\ad\ad$ \\
$\psi_2(1.6)$ \protect\cite{NeubertXi2} & $\ad-0.05\pm0.02\ad$  &
	 $0\ap\ad\ad$ \\
$\psi_3(1.6)$ \protect\cite{NeubertXi3} & $\am0.016\pm0.004$ &
	 $0.02$ \\
\hline\hline
\end{tabular}
\end{center}
\end{table}

Subleading heavy quark effects have also been calculated in the context of
a Bethe-Salpeter-inspired relativistic quark model \cite{HoldSuth}. The
results differ significantly from those here. In particular,
Ref.~\cite{HoldSuth} obtains negative values for $\psi_1$ and $\psi_3$ away
from zero-recoil, which is difficult to reconcile with both the results
here and the sum rule results.

The various predictions for subleading effects are testable. If the
complete set of form factors for $B\to D^{(*)}$ semileptonic decay were
accurately measured, one could perform an analysis similar to that of
Ref.~\cite{mySemiLep} to extract form factors similar to those in
\tabref{tab:semiLep}. Admittedly, these measurements will have to be quite
sensitive to distinguish between the models. More realistically, Neubert
\cite{NeubertReview} has shown that forward-backward asymmetry measurements
in $B\to D^*$ semileptonic decay are particularly sensitive to $\lambar$.
This provides some hope for one test of the models in the relatively near
future.

It is also interesting to see what these results tell us about the ISGW
model. \secref{sec:ISGW} demonstrated that the ISGW model does not quite
reproduce the leading-order results of heavy quark symmetry. The
discrepancy involves one form factor being multiplied by an erroneous
$2/(w+1)$. This factor varies only from $1$ to $0.77$ over the kinematic
range for $B\to D^{(*)}$ decays, however, so it does not have an
anomalously large effect. At subleading order, the dominant correction is
the current correction. The ISGW model does include the difference between
the quark mass and momentum and the meson mass and momentum, so this effect
is included.

The remaining effects calculated here are due to the heavy quark kinetic
energy and the hyperfine interaction between the quarks. The ISGW
Hamiltonian, \eqn{eq:HISGW}, does include a kinetic energy term for the
heavy quark, so this effect is also considered. It should be noted,
however, that the original ISGW calculation included the effect of the
heavy quark kinetic energy through a flavor-dependent value of $\beta$;
this does not give a consistent mass expansion at order $1/m_Q$. While the
ISGW Hamiltonian does not include spin-spin interactions, this calculation
has shown that these effects are extremely small. Non-relativistic quark
models do not seem to allow for a term of the form of $\psi_2$. Even if the
sum rule calculation of this quantity is correct, however, it does not play
a crucial role compared to subleading effects other than $\psi_3$.

By their own admission, ISGW do not include radiative QCD effects. These
effects are suppressed by $\alpha_s(\mu)/\pi \approx 10\%$, where $\mu$ is a
scale somewhere between $m_c$ and $m_b$. This is approximately the same as
$\lambar/(2m_c)$, so these effects can be as or even more important than
the $1/m_c$ effects. They should be included if the $1/m_c$ effects are
included. Although it shows some rough spots, the ISGW model is a
reasonable approximation from the vantage point of HQET.

\section{Conclusions}

The ISGW model can be simply extended to reproduce the leading-order
results of heavy quark effective theory. Using the model to estimate the
effects that occur at subleading order in the effective theory, the
functions $\psi_+$ and $\psi_2$ vanish, while the functions $\psi_1$ and
$\psi_3$ are easily calculable. The constraints on $\psi_1$ and $\psi_3$
\cite{LukeThm} are simply understood as consequences of time-independent
perturbation theory.\footnote{These constraints have also been shown
\protect\cite{BoydBrahm} to follow from the Ademollo-Gatto Theorem
\protect\cite{AdGatt}.} The calculated values of the functions are small
and positive, which is in qualitative, if not quantitative, agreement with
other estimates. These results indicate that the overall corrections to the
heavy quark limit in $B\to D^{(*)}$ semileptonic decay should be small.

\section*{Acknowledgments}

I would like to thank G. Boyd, J. Jungman and J. Rosner for many helpful
discussions. This work was supported in part by the United States
Department of Energy under Grant No.\ DE~AC02~90ER40560.

\newpage


\end{document}